\documentstyle[12pt]{article}
\def\D{{\rm d}}    
\def\e{{\rm e}}    
\begin{document}
\noindent
ISSN 0284 - 2769 \hfill{TSL/ISV-94-0092}

\hfill{February 1994}
\vskip1cm
\begin{center}
{\large \bf RADIATIVE CORRECTION SCHEMES\\IN DEEP INELASTIC MUON SCATTERING}
\vskip0.5cm
{\large B. Bade\l{}ek$^1$, D. Bardin$^2$, K. Kurek$^3$} and {\large
C. Scholz$^4$}
\end{center}

\vskip0.5cm
\noindent
$^1$ {\it Department of Physics, Uppsala University, P.O.Box 530,
751 21 Uppsala, Sweden} and

\noindent
\hspace{2mm}
{\it Institute of Experimental Physics, Warsaw University, Ho\.za 69, 00-681
Warsaw, Poland}\\
$^2$ {\it Theory Division, CERN, 1211 Geneva 23, Switzerland}  and

\noindent
\hspace{2mm}
{\it Joint Institute for Nuclear Research, 141 980 Dubna, Moscow Region,
Russia} \\
$^3$ {\it Institute of Physics, Warsaw University Branch, Lipowa 41,
15-424 Bia\l{}ystok, Poland}\\
$^4$ {\it Max Planck Institute of Nuclear Physics,
P.O.Box 103980, 69029 Heidelberg, Germany}\\

\medskip\medskip\medskip
\vskip1cm
\begin{abstract}
A description and a detailed comparison of the Mo and Tsai and the Dubna
radiative correction schemes is presented. Numerical comparisons made in the
kinematical region of the NMC high energy deep inelastic
electroproduction experiment are discussed. An overall agreement
between the two approaches in the region of low $x$ and high $y$, where
the radiative corrections are largest, is better than 2$\%$.
\end{abstract}
\medskip \medskip \medskip
\section{Introduction}
\medskip \medskip
It is well known that information on the nucleon internal structure is
contained in the electromagnetic structure functions or in the one photon
exchange cross section, $\sigma _{1\gamma}$, for the deep inelastic
lepton scattering off nucleons. However, a determination of the one photon
cross section from the data, which is a goal of electroproduction
(i.\,e.\ electron-- and muon scattering)
experiments, demands excluding contributions from other
electroweak processes. These processes account for a large fraction
of the measured cross section, especially in the low $x$ and high $y$ region.
They cannot be discarded from the measured cross section
on the event--by--event basis; the measured differential cross section
can instead be multiplied by a correction factor calculated theoretically. This
is called a radiative correction procedure.

\medskip
 The first
radiative correction scheme was created in the sixties by L.W. Mo
and Y.S. Tsai \cite{Tsai60,MTa,MTb} (MT scheme) in connection with the early
SLAC electron scattering experiments. Another approach, originally formulated
for the CERN muon scattering experiment planned by the BCDMS Collaboration,
was suggested by the
Dubna group in the seventies \cite{Dubna70} and upgraded
later \cite{Dubna80}(D scheme). In the analysis of the data from the
deep inelastic electroproduction experiments, both schemes
were extensively used  (cf.\ e.\,g.~\cite{TS:PR}), the MT scheme often in the
exact (i.\,e.\ no `peaking approximation') and upgraded version.
However, the precision of the recent
experiments is so high, see \ e.\,g.\cite{NMCF2}, that also
the radiative correction
procedure has to be based on more precise theoretical calculations.
 Therefore
understanding similarities and differences in the two radiatiative
correction schemes is of ultimate
importance for concluding about consistency of results coming from different
experiments.

\medskip
The goal of this paper is to compare analytically as well as numerically
the most upgraded versions of the two approaches. Our experience
with the radiative correction procedure in the deep inelastic experiments
carried out at CERN by the EMC, NMC and BCDMS
Collaborations
allows to point out the problems encountered during the application of
this procedure in the data analysis.
The previous comparisons between the two considered methods were limited to
analysing the numerical results of the early versions of the schemes
\cite{TS} or only dealt with a subset of the radiative processes
\cite{konkurencja}.

\medskip
The paper is organized
as follows. In Section 2 the deep inelastic kinematics and cross sections are
 defined. Sections 3 and 4, together with Appendices A and B contain
a description of the MT and D schemes respectively.
A complete set of formulae is given for each scheme
using the original notation. Only minor simplifications and changes
are introduced for the sake of clarity.
Useful relations between the respective notations are given
in the Appendix C. The schemes are then compared in Sections 5 (theoretical
ideas) and 6 (numerical results) and finally a summary is given in Section 7.

\medskip\medskip\medskip
\section{Basic Definitions and Kinematics of Deep Inelastic Scattering}
\medskip\medskip
As mentioned above, a goal of electroproduction experiments is to extract
the differential cross section in the one photon exchange
approximation (fig.~1a)
from the data. This cross  section can be expressed in the following
way by the structure functions $F_1(x,Q^2)$ and $F_2(x,Q^2)$ of the target:
\begin{eqnarray}
{\D^2\sigma _{1\gamma}(x,Q^2)\over \D Q^2\D x}={4\pi \alpha^2\over
 Q^4}\left[\left(1-y-
{Mxy\over 2E}\right) {F_2(x,Q^2)\over x}+\left (1-{2m^2 \over Q^2}\right )
y^2F_1(x,Q^2)\right]\,.
\label{basic}
\end{eqnarray}
In this equation $\alpha$ is the fine structure constant, $m$ is the electron
(muon) mass, $Q^2 = -q^2$ where $q^2$ is the square of the four--momentum
transfer between the incoming and outgoing lepton, $x = Q^2$/$(2M\nu)$
the Bjorken scaling variable, $M$ is taken
as the proton mass, $E$ and $\nu$ are the lepton's incident energy
and energy transfer in the proton rest frame
and $y=\nu/E$.

\medskip
The one photon exchange process described by the eq.~(\ref{basic}), is a part
of the lowest order (or Born) electroproduction cross section, $\sigma ^B$.
The other part of $\sigma ^B$ proceeds from the $Z^0$ boson
exchange.  The two contributions cannot be separated
experimentally. However in the present fixed target experiments the involved
virtualities (i.\,e.\ the $Q^2$ values) are small comparing to the $Z^0$ mass
squared and therefore eq.(\ref{basic}) is a good approximation of the Born
cross section. The $Z^0$ contribution will be discussed later in more detail.

\medskip
The differential cross section (1) can also be expressed in terms
of structure functions $F_2(x,Q^2)$ and $R(x,Q^2)$:
\begin{eqnarray}
{\D^2\sigma _{1\gamma}(x,Q^2)\over \D Q^2\D x}=
{4\pi \alpha^2\over Q^4}{F_2\over x}
\left[1-y-
{Mxy\over 2E} + \left (1-{2m^2\over Q^2}\right )
{y^2(1+4M^2x^2/Q^2)\over 2(1+R)}\right]\ ,
\label{basicp}
\end{eqnarray}
where $R$ is defined as:
\begin{equation}
R(x,Q^2)={\sigma_L\over\sigma_T}={(1+4M^2x^2/Q^2)F_2\over 2xF_1}-1;
\label{r}
\end{equation}
$\sigma_L$ and $\sigma_T$ denote the cross sections
for the longitudinally and transversally polarised virtual photon
respectively.

\medskip
As mentioned earlier the radiative events account for
a large
fraction of the measured cross section, $\sigma _{\rm meas}$.
These effects may lead to a wrong interpretation of the measured event
kinematics. For example
an elastic scattering from the target can be mistaken for a deep inelastic
event if it is accompanied by an energetic bremsstrahlung photon not
measured in the experiment.
The magnitude of the radiative effects in the measured cross
section will be characterized by the so called radiative correction
factor, $\eta (x,y)$, defined as follows:
\begin{equation}
\eta (x,y) = {\sigma _{1\gamma }\over \sigma_{\rm meas}}\,.
\label{eta}
\end{equation}
This factor is used in the data analysis \cite{NMCF2} and will also be employed
in comparing results of calculations between different radiative correction
schemes.

\medskip
Finally we have to stress that in the most of the deep inelastic experiments
only the inclusive measurements are performed. This means that only
incident and scattered leptons are measured and kinematics of the reaction
is defined by the leptonic observables. However the kinematics of radiative
events
cannot be defined by leptonic variables only, e.\,g.\ a bremsstrahlung photon
emission by the lepton results in a {\it measured} $Q^2$ different from
the {\it actual} one. Therefore, when appropriate,
we shall make a clear distinction between the hadron-- and lepton--defined
variables.

\medskip\medskip\medskip
\section{Mo and Tsai Scheme}
\medskip\medskip
In the original MT scheme \cite{Tsai60,MTa,MTb} the following
processes were considered to contribute
to the measured electron--proton inelastic cross section\footnote
{The original MT scheme was formulated for the electron scattering;
therefore a large effort was put to quantify the effects of the energy loss
in the electron passage through the target. We shall neglect them in this
paper since in the examples of practical applications we shall deal with
muon scattering only.}: real photon
bremsstrahlung from an initial and final electron (fig.~1b),
vertex correction
(fig.~1c) and vacuum polarisation correction (fig.~1d). In the latter one only
the electron and muon loops were originally considered. The proton
structure was accounted for through the two unspecified structure functions.
The MT scheme formulation is thus model independent.
The mathematical formulation of the MT is non-covariant.

\medskip
Corrections depicted by the diagrams in fig.~1 can be divided into two
groups: emission of real photons (fig.~1b) with energy larger than $\Delta$
and those with energy smaller than $\Delta$ (fig.~1b) together with virtual
corrections (fig.~1c,d).
Contributions from the soft photon emission and from the vertex correction
{\it separately} are infrared divergent but the divergences cancel when
the contributions are considered jointly \cite{BD}.
The parameter $\Delta$ may have a meaning of
the energy resolution or in another words a maximal energy of the emitted
photon which is still not detectable in the experiment.
It is often called the `infrared cut--off' parameter.
This interpretation implies that $\Delta$ may not be too large.
Too small $\Delta$ must also be avoided since it
may cause numerical instabilities in the computation of the soft photon
contribution.
The numerical results of the calculations should not depend on $\Delta$.

\medskip
In the MT approach the measured cross section can be expressed as follows
(see Appendix A for the definitions of variables used in MT publications):
\begin{eqnarray}
{\D^2\sigma_{\rm meas}\over \D\nu \D\Omega} &=& \e^{-\delta_R(\Delta)} F(Q^2)
{\D^2\sigma_{1\gamma}\over \D\nu \D\Omega}
 + {\D^2\sigma_{\rm tails}\over \D\nu \D\Omega} ,
\label{mtmaster}
\end{eqnarray}
where
\begin{eqnarray}
\delta_R(\Delta) = {\alpha \over \pi}\left(\ln{E_s\over \Delta} + \ln{E_p\over
\Delta}\right)\left(\ln{Q^2\over m^2} - 1\right)
\label{mtdel}
\end{eqnarray}
is a residuum of the cancellation of the infrared divergent terms and takes
into account all soft photon emissions in the lowest order of $\alpha$.
It is a well known fact that the infrared divergencies cancel in each
order in
$\alpha$ \cite {BD} and therefore it is possible to sum up contributions from
all soft photon emissions \cite{YFS}.
The exponential factor, $\e^{-\delta_R(\Delta)}$, in eq.(\ref{mtmaster}) is
a result of this summation. The function $F(Q^2)$ contains all
$\Delta-$independent terms:
\begin{eqnarray}
F(Q^2) = 1 + \delta_{\rm vac}^e +
\delta_{\rm vac}^{\mu} + \delta_{\rm vtx} + \delta_s ,
\label{mtf}
\end{eqnarray}
where\footnote{
$\delta^{e,\mu}_{\rm vac}$ given here is the full formula. The approximation
${2\alpha/\pi}(-{5/ 9}
+ {1/3}\ln{Q^2/m_{e,\mu}^2})$ holds for $Q^2\gg m_{e,\mu}^2$
(see e.\,g.\ eqs~5 and A1 from ref.~\cite{Tsai60}). For extremely low $Q^2$
formula (8) converges to $Q^2/15m_{e,\mu}^2$.}
\begin{eqnarray}
\delta_{\rm vac}^{e,\mu } &=& \frac{2\alpha}{\pi}\left[\frac{-5}{9}+
   \frac{4m_{e,\mu}^2}{3Q^2}+\frac{1}{3}\sqrt{1+\frac{4m_{e,\mu}^2}{Q^2}}
  \cdot  \left(1-\frac{2m_{e,\mu}^2}{Q^2}\right) \ln \left(
        \frac{\sqrt{1+4m_{e,\mu}^2/Q^2}+1}{\sqrt{1+4m_{e,\mu}^2/Q^2}-1}
           \right) \right]  \\
\label{mtvac}
\delta_{\rm vtx}&=&{2\alpha\over \pi}\left(-1
+ {3\over 4}\ln{Q^2\over m^2}\right)  \\
\label{mtvtx}
\delta_s&=&{\alpha\over \pi}\left[{1\over 6}\pi^2 - \Phi \left(\cos^2{\theta
\over 2}\right) + \Phi \left({E_p-E_s\over E_p}\right) +
                  \Phi \left({E_s-E_p\over E_s}\right)\right]
\label{mts}
\end{eqnarray}
\noindent
and $\Phi$ the Spence function\footnote{In the original MT formulation
\cite{MTb} the `$q^2$' symbol was used in eqs (\ref{mtdel}) - (\ref{mtvac}).
This variable, however, was then redefined to become
the {\it measured} four momentum transfer, $q^2 = (s-p)^2$, and thus coincides
with our definition of $-Q^2$. The latter was thus used here for clarity.}.
The second term in eq.(\ref{mtmaster}), $\sigma_{\rm tails}$, accounts for
contribution from processes where the real photons of energy larger than
$\Delta$ are emitted:
\begin{equation}
{\D^2\sigma_{\rm tails}\over \D\nu \D\Omega} =
{\D^2\sigma(\omega > \Delta)\over \D\nu
 \D\Omega}
= \int^{M^{\rm max}_j}_{M_j=M}\D M_j~{\D^2\sigma_{j,r}\over \D\nu
\D\Omega}\,,\\
\label{mtails}
\end{equation}
where $M$ denotes a target mass, $M^{\rm max}_j=\sqrt{M^2-Q^2+2M(\nu-
\Delta)}$ and $\sigma_{j,r}$ is given by the formula (A.24) in \cite{MTb} and
quoted in Appendix A; $M_j$ is an effective mass of the hadronic
final state. The integration over $M_j$ in eq.(\ref{mtails}) means that all
the final hadronic states contribute to the cross section measured in
a kinematical point $(Q^2,\nu)$, fig.~2, so called `radiative tails':
elastic ($M_j = M$), resonance production ($M_j = M_{res}$) ... and
deep inelastic tails, i.\,e.\ tails from the continuum.

\medskip
Equation (\ref{mtmaster}) does not take into account double-- (multi--) photon
exchange
reactions (fig.~1e) nor any radiative correction from the hadron current
(fig.~1f,g).
These effects were estimated only for the {\it elastic} $e-p$ interaction
\cite{MTa}. Application of these calculations to the inelastic interaction
is incorrect. Thus these results were discarded in practical
applications
of the MT scheme. Evaluation of these effects in the inelastic $e-p$
interactions was not done in this scheme. To do this a model
of the proton internal structure is necessary. The best framework
presently is the quark parton model (QPM).

\medskip
In the upgraded version of the MT scheme, used in the analysis of the NMC
results \cite{n/plong}, the $\tau^+\tau^-$ and $q\bar q$ loops in the vacuum
polarisation
\cite{BD,jeger} and the virtual photon $- Z^0$ boson interference \cite{TS:PR}
were also included.

\medskip\medskip\medskip
\section{Dubna Scheme}
\medskip\medskip
In the Dubna scheme the calculations of the deep inelastic processes
are based on a mixed approach which uses both a model independent
and the quark parton model treatment of the radiative corrections.
The radiative corrections to the leptonic current (fig.~1b,c) are
calculated within the same model independent approach as
in the Mo and Tsai scheme except the way of treating the soft bremsstrahlung
photons. All other corrections are calculated using the
QPM approach.

\medskip
In the inclusive electroproduction experiments the radiative and non--radiative
events cannot be distinguished and therefore the
$\sigma_{\rm meas}$ should not depend on any parameter $\Delta$,
a property, which {\it implicitly} holds in Mo and Tsai scheme.
The D scheme is {\it explicitly} $\Delta$ independent as a result
of an integration over the whole bremsstrahlung photon phase space.
The relevant procedure is described in detail in \cite{teupitz}.

\medskip
Applying the QPM and the fact that quarks are point--like
objects permit in principle to calculate all QED processes
to all orders of $\alpha$. This means that
in addition to the diagrams in fig.~1.a-d
also the hadron current corrections (fig.~1.e-g) can be
evaluated. The diagrams of fig.~1.e-g thus become as in fig.~3a-c.
All these processes were taken into account as well as the
lowest order weak corrections, not shown here, which can also be calculated
in the framework of the quark-parton model.
The outline of the radiative corrections calculations in the D scheme
is given below; the details are given in refs \cite{teupitz,disep}.

\medskip
The measured cross section is now expressed as follows%
\footnote{Additional factor $\alpha /\pi$ in the third row of
eq.~(\ref{master})
is hidden in the definitions of the $R$ functions.}:
\begin{eqnarray}
 {\D^2\sigma_{\rm meas}\over \D Q^2\D x}
 &=& {\D^2\sigma ^B\over \D Q^2\D x}
 \Biggl\{\e^{-\delta_R(x,Q^2)} +
 \delta^{VR}(x,Q^2) \Biggr\}  \nonumber  \\
 &+&{\D^2\sigma_{\rm in. tail}\over \D Q^2\D x}-
 {\D^2\sigma^{IR}\over \D Q^2\D x}  \nonumber  \\
 &+& {2\pi\alpha^2\over Q^4}
    \sum_{B=\gamma,I,Z} \sum_{b=i,q}~ \sum_{Q,\bar Q}
    c_b K(B,p) \left[ V(B,p)R^V_b(B)+pA(B,p)R^A_b(B) \right]
\nonumber   \\
 &+& {\D^2\sigma_{\rm el. tails} \over \D Q^2\D x}.
\label{master}
\end{eqnarray}
The first two rows of this formula represent the results of the model
independent calculations of the radiative corrections to the leptonic current.
The third row represents the quark parton model description of the
lepton--hadron interactions (fig.~3a) and of the radiative corrections to the
hadronic current (fig.~3b,c) as well as certain interference terms.
Finally the last term in eq.~(\ref{master}),
${\D^2\sigma_{\rm el. tails}/\D Q^2\D x}$, describes
the elastic and resonance radiative `tails'.

\medskip
The ${\D^2\sigma ^B/\D Q^2\D x}$ cross section in eq.~(\ref{master}) denotes
the
full Born cross section for the deep inelastic scattering, i.\,e.\ the cross
section containing both the one photon-- and one $Z^0$ boson exchange
contributions:
\begin{equation}
 {\D^2\sigma^{B}\over \D Q^2\D x}
 = {2\pi \alpha^2 y \over Sx}  \sum_{i=1}^{3}
     {\cal A}_{i}(x,Q^2) {1 \over Q^4} {\cal S}_{i}^{B}(y,Q^2) ,
\label{born}      \\
\end{equation}

\noindent
where the functions ${\cal S}_{i}^{B}(y,Q^2)$ are:

\begin{eqnarray}
{\cal S}_{1}^{B}(y,Q^2) &=& Q^2-2m^2,  \nonumber  \\
{\cal S}_{2}^{B}(y,Q^2) &=& 2[(1-y)S^2-M^{2}Q^{2}], \nonumber \\
{\cal S}_{3}^{B}(y,Q^2) &=& 2Q^{2}S(2-y).  \nonumber \\
\label{rad}
\end{eqnarray}
\noindent
with $Q^2=Sxy$ and $S=2ME$. The functions ${\cal A}_{i}$ are given in
Appendix B.

\medskip
The $\delta _R$ in eq.~(\ref{master}) is responsible for those parts of the
{\it soft} and {\it hard collinear} photon emissions which could
be resummed to all orders using the covariant exponentiation procedure,
\cite{YFS,shumeiko}. It reads:
\begin{equation}
\delta_{R}=-{\alpha\over{\pi}}\left(\ln {Q^2 \over m^2}-1\right)
\ln{y^2(1-x)^2 \over (1-yx)(1-y(1-x))}.
\label{delinf}
\end{equation}
\noindent
The $\delta^{VR}$ correction factor in eq.(\ref{master}) is a remnant
of the exponentiation and of the subtraction procedure used to disentangle
the infrared divergent terms from the ${\D^2\sigma_{\rm in. tail}/
\D Q^2\D x}$ cross section \cite{teupitz}, see below. It thus contains
the vertex correction, fig.~1.c and is given by
\begin{equation}
\delta^{VR}=\delta_{\rm vtx}-{\alpha\over{2 \pi}}
\ln^2{(1-yx) \over (1-y(1-x))} +
\Phi\left[{(1-y) \over (1-yx)(1-y(1-x))} \right] - \Phi(1).
\label{delrem}
\end{equation}

\medskip
The ${\D^2\sigma_{\rm in. tail}/\D Q^2\D x}$ in eq.~(\ref{master}) describes
the
inelastic radiative tail for the {\it lepton current correction} only:
\begin{equation}
{\D^2\sigma_{\rm in. tail}\over \D Q^2\D x}
={2\alpha^3 y \over Sx}
    \int \!\!\int \D Q^2_h  \D M^2_h \sum_{i=1}^{3}
    {\cal A}_{i}(x_h,Q_h^2) {1 \over Q^4_h}
    {\cal S}_{i}(y,Q^2,y_h,Q^2_h).
\label{dtails}
\end{equation}
In this formula $M_h$ is the invariant mass of the final
hadronic system  and the variables bearing a subscript
 `$h$' refer to virtual photon--target
vertex in contrast to the variables measured in the inclusive
electroproduction experiment. For their definitions see Appendix B where also
the explicit expressions for the radiator functions ${\cal S}_{i}$ are given.

\medskip
It is a well known fact that the
${\D^2\sigma_{\rm in. tail}/ \D Q^2\D x}$ cross section
is infrared divergent. To regularize it a simple trick (`fixation
procedure') of adding and subtracting an extra term
\begin{equation}
{\D^2\sigma^{IR}\over \D Q^2\D x}
= {\D^2\sigma^{B}\over \D Q^2\D x}
    \int\!\! \int \D Q^2_h \D M^2_h
    {\cal F}^{IR}(y,Q^2,y_h,Q^2_h),
\label{dir}
\end{equation}
to ${\D^2\sigma_{\rm in.
tail}/ \D Q^2\D x}$ was employed \cite{YFS,teupitz}. In the added term
an integration over a full photon phase space was carried out,
resulting in the above given expressions for $\delta _R$ and $\delta ^{VR}$.
The subtracted term appears explicitly in the second row of eq.(\ref{master}),
so that the difference $\D^2\sigma_{\rm in. tail}/ \D Q^2\D x -
{\D^2\sigma^{IR}/ \D Q^2\D x}$  in eq.~(\ref{master}) is finite
over the full
kinematic domain of $Q^2_h$ and $M_h^2$. This method is a key point of the
D scheme, making it {\it explicitely} $\Delta$ independent.
The function ${\cal F}^{IR}$ is given in Appendix B.

\medskip
The third row in the eq.~(\ref{master}) represents the quark parton
model calculations and contains clearly visible vector ($\gamma$ and $Z^0$)
and axial (only $Z^0$) contributions.
Index $B$ runs over the photon exchange ($\gamma$), the $Z^0$ boson
exchange ($Z$) and their interference ($I$). Index $b$ stands for the
scattering with the single photon emission by the quark ($q$), fig.~3c, and its
interference with the photon emission by the lepton ($i$). The double photon
exchange (fig.~3a) as well as the vertex corrections on the quark line
(fig.~3b)
are also hidden there. Both scattering off quarks
($Q$) and antiquarks ($\bar Q$) were considered \footnote{In the original
formulation \cite{disep} the index $b$ assumes also values 0 and l which
correspond to the contributions from processes in fig.~1a,b,c.
Here they are included in the model independent parts of eq.~(\ref{master})}.
Coefficients $c_b$ are equal to $Q^2_Q$ or $Q_{\mu}Q_Q$ for the process of
bremsstrahlung photon emission from the quark or for the interference term
between the photon emission from the lepton and from the quark respectively;
$Q_\mu$ and $Q_Q$ are the charges of the lepton and of the quark given
in the electron charge units. The sign factor $p$ is defined as: $p=p_\mu p_Q$
where $p_{\mu ,Q}=\pm$ 1 for particle (antiparticle).
For a detailed form of the coupling strength factors
$K(B,p)$, modified vector, $V(B,p)$ and axial, $A(B,p)$ couplings
as well as of the functions $R_b^{V,A}(B)$ we refer the reader to ref.
\cite{disep}.

\medskip
The problem of the soft photons' emission from the quarks can be
uniquely
solved since the quarks are not observed and thus the quark states are all
summed up in the cross section. The infrared divergencies for initial--
and final state quarks as well as
the quark mass singularities for the final state quarks
cancel in each order in $\alpha$ \cite{BD,disep}. Mass singularities
associated with initial quark are included in the definition of the proton
structure function.

\medskip
The last row in eq.~(\ref{master}), ${\D^2\sigma_{\rm el. tails}/ \D Q^2\D x}$,
is in the D scheme treated essentially in the same way as in the MT
except that it is formulated in a covariant way.
Finally the vacuum polarisation, fig.~1d, was taken into
account via the `running' $\alpha(Q^2)$ which in the $Q^2 \gg m_f^2$
approximation ($m_f$ stands for the lepton and quark effective masses),
\cite{jeger} is defined as follows:

\begin{equation}
 \alpha(Q^2)=\frac{\alpha}{1-{1\over 2}\sum_{f}c_f Q_f^2
\delta_{\rm vac}^f} ,
\label{alfa}
\end{equation}
where $c_{f}$ and $Q_{f}$ are the colour factor and the electric charge of
fermions $f~ (c_{f} = Q_{f} = 1$ for leptons); `$f$' runs over all leptons
and quarks.

\medskip
We shall close this section with the following remarks:
first, although not shown explicitely in eq.~(\ref{master}),
the weak loop correction contribution is also present in the D formulation.
It is calculated within the QPM framework; details are given in \cite{disep}.
Second, the ${\cal O}(\alpha^{2})$ corrections
($\alpha^4$ contributions to the cross section) not shown explicitely,
 were also implemented. For the elastic radiative tail they were calculated
completely in the first paper of ref. \cite{Dubna80}, while for the inelastic
continuum they were implemented in an approximate way, described in
the second paper of ref. \cite{Dubna80}.

\medskip\medskip\medskip
\section{Comparison of the MT and D Schemes}
\medskip\medskip
Results of deep inelastic experiments were analysed using
either MT or D radiative correction schemes. Therefore it is of ultimate
importance to understand the differences and similarities between the two
approaches.
In this section we shall make a brief summary of theoretical ideas
in the two schemes; numerical comparison will be presented in the next section.
To faciliate the comparison, the relations between the variables used in
the MT and D formulae are given in Appendix C.

\medskip
The D scheme is formulated in a covariant way and its model dependent part
is based on the quark parton model. The covariant formulation means that
all formulae are expressed in terms of the Lorentz invariants and are thus
independent of the choice of the reference frame. This also means
that the D scheme is explicitly independent of the infrared cut--off
parameter $\Delta$, cf.\ eq.~(\ref{master}). The $\Delta-$
independence was obtained through a special mathematical procedure.

\medskip
The MT scheme is not covariantly formulated. It should also be
$\Delta-$independent which formally means that the
derivative over $\Delta$ of the right hand side of
eq.(\ref{mtmaster}) should be equal to zero. The calculations show that this
does not hold and that a certain dependence of the results on $\Delta$ should
be
observed. For not too small $\Delta$ this dependence is very weak.
For very small $\Delta$ (apart of numerical instabilities),
eq.(\ref{mtmaster}) depends much stronger on the value of $\Delta$ which in
this case should
be calculated separately for every kinematical point (i.e. for every $(x,y)$
value) in order to minimise the dependence.

\medskip
The elastic and resonance tails (i.\,e.\ the contribution of the reactions
with the elastic and resonance final states) are calculated in the same way
in both schemes. The inelastic tails (the contribution
of the reactions with the inelastic final states) originating from the
leptonic bremsstrahlung are treated differently but the differences are
 purely mathematical.
The inelastic tail in the MT is calculated with the same formula
as the elastic one (eq.~(\ref{mtails})) where the inelasticity of the process
is taken into account
through integration over kinematically allowed final state masses.
The corresponding structure functions, $W^j_{1,2}$(see Appendix A),
are not specified;
in practical applications the electromagnetic structure functions are used.
The leptonic inelastic tail in D is also calculated with arbitrary structure
functions while tails originating from the hadronic bremsstrahlung
were calculated within the quark parton model framework.

\medskip
The usage of the parton model in the D
scheme allows to calculate the hadron current corrections in the inelastic case
as well as a double photon exchange process; they are calculated
up to $\alpha^3$. The hadron current corrections cannot be calculated
in the MT approach and therefore such corrections were not taken into account.
The $q\bar q$ loop is also naturally present in the vacuum polarisation
process in the D scheme while in the MT it was only added later. The same
is true for the weak interactions contributions ($Z^0$ exchange and the
$\gamma -Z^0$ interference). The weak contributions are
however very small as compared to the present experimental resolution.

\medskip
Finally the $\alpha^4$ lepton current corrections were partially taken
into account in the D scheme but not in the MT.

\medskip\medskip\medskip
\section{Numerical Calculations in the MT and D schemes}
\medskip\medskip
A tremendous increase of the accuracy of deep inelastic electroproduction
experiments demands a similar increase of the accuracy of the radiative
corrections calculations. Therefore the early versions of these calculations
have constantly been improved. At the same time comparisons between
the two considered schemes were made, \cite{TS:PR,TS,konkurencja}.
In the analysis of the
recent very precise mesurements by the NMC \cite{NMCF2,n/plong} both schemes
were used in their most upgraded versions and applied for nucleon and nuclear
targets.
 The MT code, named FERRAD35,
apart of the processes contained in the eqs (\ref{mtmaster})--(\ref{mtails})
 included also the tau lepton--
and quark loops and the photon$-Z^0$-boson
interference. The tails were treated in an exact
way in contrast to the peaking approximation \cite{MTa,MTb} applicable
for the electron deep inelastic scattering,
A detailed input information (structure functions, form factors,
nuclear structure models, etc) was introduced \cite{PhD}.
The D code, named TERAD86, was in principle used in the same version as
sketched by the formulae (\ref{master}) -- (\ref{alfa}) and employed in the
dedicated BCDMS experiment \cite{TS:PR}, except that the FERRAD35 input
information was supplied. In this section we shall compare the FERRAD35
and TERAD86 results on the radiative
correction factors, $\eta$, as well as on the elastic radiative tails.
Comparisons will be done in the kinematical region of the NMC positive
muon--proton deep inelastic scattering experiment, i.e. for 0.003$< x <$0.9
and 0.1$< y <$0.9 at the incident muon energy 280 GeV.
The input information in the calculations was as follows:
Gari and Kr\"umpelman proton form factor parametrisation, \cite{protff}
structure function $F_2$ as measured and parametrised by the NMC, \cite{NMCF2},
and finally for the $R(x,Q^2)$ the parametrisation of SLAC was taken for all
$x$ and $Q^2 > $0.35 GeV$^2$. For smaller $Q^2$ the value of $R$ was assumed
to be constant and equal to the value at $Q^2 =$ 0.35 GeV$^2$. The proton
resonances were neglected. For further details see refs \cite{n/plong} and
\cite{PhD}.

\medskip
The main numerical problem in the radiative correction programs is an
integration of the radiative tails. For example, the integrand
of eq.~(\ref{mtails}) changes by 27 orders of magnitude within the integration
interval. It is obvious that this function should be
integrated either by a high accuracy routines (e.\,g.\ the CERNLIB GAUSS
routine demands an accuracy parameter $\varepsilon <$ $10^{-12}$) or by
dividing the integration interval into many small sections. It seems however
that the best method would be to change the integration variable from
$\cos\theta_k$ to log\,$(-q^2)$ but also in this case
high accuracy of the integrating routines is needed.
This method was used in TERAD86 while in FERRAD35 dividing the
$\cos\theta_k$ integration interval into many sections was normally
employed (logarithmic
integration in FERRAD35 was also tried and gave the same results).
Due to the subtraction procedure in D the integrands in eqs~(\ref{dtails})
and (\ref{dir})
are fairly smooth and do not demand any extreme precision of integration.

\medskip
The dependence of the FERRAD35 results on the parameter $\Delta $ was carefully
studied. The results are presented in fig.~4. In the region of low $x$
and large $y$ where the radiative correction factor $\eta $ is largest,
the results are only weakly dependent on $\Delta $ for $\Delta >$200 MeV.
Results given below were obtained with $\Delta =$ 280 MeV.

\medskip
Comparison of the FERRAD35 and TERAD86 results is presented in figs~ 5--6.
Radiative corrections are very large, exceeding 50$\,\%$ at low $x$ and high
$y$, cf.\ fig.~5. In that region the agreement between the results of the two
schemes
is better than 2$\%$, cf. fig.6 (closed symbols). However the $\tau\bar\tau$
and $q\bar q$ loop contributions in the vacuum polarisation process (fig.1.d),
absent in the original version of the MT scheme and included in FERRAD35, give
up to 2$\%$ contribution to the radiative correction factor in most of
the kinematic region (fig.6, open symbols denote results of calculations
{\it without} those contributions).
Fluctuations visible in the high $x$ part of the curves in fig.6 come from
numerical instabilities of FERRAD35 in that region.

\medskip\medskip\medskip
\section{Summary}
The two existing schemes of radiative correction
procedure, the Mo and Tsai and the Dubna ones are differently formulated
and are (partially) based on a different physics approaches. Both were
extensively used in analysing the high energy experimental data. In this paper
we presented the two schemes in detail and compared them analytically
and numerically from the point of view of their effect on the results of
the deep inelastic positive muon scattering from a proton target at 280 GeV.
To this aim we used the
latest version of the D and the upgraded version of the MT programs.
The latter included the $\tau ^+\tau ^-$ and $q\bar q$
loops in the vacuum polarisation and the virtual photon$-Z^0$ boson
interference terms, all absent in the original formulation.
In contrast to the $\gamma-Z^0$ interference the quark loop contribution
turned out to be quite substantial, changing the total radiative correction
by about 2$\%$ in the measured region.

\medskip
The MT scheme contains the `infrared cut--off' parameter, $\Delta$.
The results should not depend on its value (provided it is not too large
and not too low) and indeed it is approximately so when $\Delta$ is
equal to about 0.1$\%$ of the beam energy value. The covariant
formulation of the D scheme excludes the existence of such parameter.

\medskip
The overall radiative correction reaches 50$\%$ at low $x$ and high $y$.
Calculated from the two schemes the corrections agree to better than 2$\%$
in this region. Differences are thus insignificant over the most of the
phase space covered in the fixed target DIS experiments.
They are of the order of other systematic errors in the data analysis
\cite{NMCF2}.

\medskip
Neither of the two radiative correction schemes contain a contribution
from the multiphoton exchange process to the {\it elastic} radiative tail
which may be important for heavy nuclear targets. Results of quantitative
estimates of those processes, relevant for the heavy target data currently
analysed by the NMC, are discussed in a separate paper \cite{multig}.

\medskip\medskip\medskip
\noindent
{\Large {\bf {Acknowledgements}}}

\medskip\medskip\medskip
We thank our colleagues from the EMC and NMC for never--ending discussions
of radiative corrections and for the enjoyable research collaboration.
We are indebted to A. Akhundov for critical reading of the manuscript and
important comments. DB is very much obliged to L. Kalinovskaya and T. Riemann
for valuable discussions.
This research was supported in part by the Polish Committee for Scientific
Research, grant number  2 P302 069 04 and by Bundesministerium f\"ur
Forschung und Technologie.

\medskip\medskip\medskip
\section{Appendix A}
\medskip\medskip
Below we summarize variables used in
the MT formulation. The metric used is such that $ps =E_pE_s - \bar p\bar s$
and
four vector components are in the laboratory system.
Notation is explained in fig.~7 and the coordinate system is that of fig.~8.

\noindent
\begin{tabbing}
abc\=$\alpha$qwertyuiopasdfghjklaaaarttttttt\=fine structure constant \kill
   \>$s=(E_s,\overline s)$              \>four momentum of the incident lepton
\\
   \>$p=(E_p,\overline p)$              \>four momentum of the scattered lepton
\\
   \>$\theta(\Omega)$           \>lepton scattering angle (solid angle), \\
                        \>      \>$\cos\theta = \overline s\overline p/\mid
\overline s\mid \mid \overline p\mid$ \\
   \>$\theta_k, \phi_k$         \>bremsstrahlung photon emission angles \\
   \>$\theta_v, \phi_v$         \>virtual photon emission angles \\
   \>$t = (M,\overline 0)$              \>four momentum of the target proton \\
   \>$k = (\omega ,\overline k)$        \>four momentum of the bremsstrahlung
photon \\
   \>$p_f = s + t - p - k$              \>four momentum of the final hadronic
system \\
   \>$q^2 = (s-p-k)^2 = (p_f-t)^2$      \>four momentum transfer \\
   \>$-Q^2 = (s - p)^2$         \>measured four momentum transfer \\
\end{tabbing}

\medskip
In the one photon exchange approximation and assuming one photon emission,
the radiative tail from the $j$'th mass level can be written as (formula
(A.24) in \cite {MTb}):

\begin{eqnarray*}
{\D^2\sigma_{j,r}\over \D\Omega \D E_p} &=&
{\alpha^3\over 2\pi}\left({E_p\over E_s}
\right)\int ^1_{-1} {2M\omega \D (\cos\theta_k)\over q^4(u_0 - \mid
\overline u\mid \cos\theta_k)}  \\
&&\left(W^j_2(q^2)\left\{ {-am^2\over x^3}\left[2E_s(E_p+\omega )
+ {q^2\over 2}\right] - {a'm^2\over y^3}\left[2E_p(E_s - \omega ) +
{q^2\over 2}\right]\right.\right.  \\
&&- 2 + 2\nu (x^{-1} - y^{-1})\left\{ m^2(s\cdot p - \omega ^2) +
(s\cdot p) \left[2E_sE_p - (s\cdot p) + \omega (E_s - E_p)\right]\right\} \\
&&+ x^{-1}\left[2\left(E_sE_p + E_s\omega + E_p^2\right) + {q^2\over 2}
- (s\cdot p) - m^2\right]  \\
&&- y^{-1}\left.\left[2\left(E_sE_p - E_p\omega + E_s^2\right) + {q^2\over 2} -
(s\cdot p) - m^2\right]\right\}  \\
&&+ W_1^j(q^2)\left[\left({a\over x^3} + {a'\over y^3}\right)
m^2(2m^2 + q^2) + 4 \right. \\
&&+ 4\nu \left.\left.\left(x^{-1} - y^{-1}\right)(s\cdot p)\left(s\cdot p -
2m^2\right) + \left(x^{-1} - y^{-1}\right)\left(2s\cdot p +
2m^2 - q^2\right)\rule{0cm}{4ex}\right]\right)
\end{eqnarray*}
\noindent

where

\begin{eqnarray*}
\omega &=& {1\over 2}\left(u^2 - M_j^2\right)/\left (u_0 - \mid\overline u\mid
\cos\theta_k\right) \\
u &=& s + t - p\: =\: p_f + k  \\
u_0 &=& E_s + M - E_p \\
\mid\overline u\mid &=& \left(u_0^2 - u^2\right)^{1/2} \\
u^2 &=& 2m^2 + M^2 - 2(s\cdot p) + 2M\left(E_s - E_p\right) \\
q^2 &=& 2m^2 - 2(s\cdot p) -2\omega \left(E_s - E_p\right) + 2\omega \mid
\overline u\mid \cos\theta_k \\
a &=& \omega \left(E_p - \mid\overline p\mid \cos\theta_p \cos\theta_k\right)
\\
a'&=& \omega \left(E_s - \mid\overline s\mid \cos\theta_s \cos\theta_k\right)
\\
b &=& -\omega\mid\overline p\mid \sin\theta_p \sin\theta_k \\
\nu &=& (a' - a)^{-1} \\
\cos\theta_p &=& (\mid\overline s\mid \cos\theta - \mid\overline p\mid)/\mid
\overline u\mid \\
\cos\theta_s &=& (\mid\overline s\mid - \mid\overline p\mid \cos\theta )/ \mid
\overline u\mid \\
x &=& \left(a^2 - b^2\right)^{1/2} \\
y &=& \left(a'^2 - b^2\right)^{1/2} \\
\end{eqnarray*}

\medskip
\noindent
$W_1$, $W_2$ denote structure functions; in particular
$W_{1,2}^j (q^2)$ are the structure functions at four momentum
transfer $q^2$ and invariant mass of the hadronic final state $M_j$.
$W_{1,2}$ are connected with the functions $F_{1,2}$ of eq.~(\ref{basic})
in the following way: $F_2 = \nu W_2$ and $F_1 = MW_1$. Observe that the
meaning of the $\nu$, $x$ and $y$
wariables used in the MT formulation is different from their generally accepted
meaning as the DIS variables.

\noindent
\medskip\medskip\medskip
\section {Appendix B}
\medskip \medskip
Below the exact expressions for certain functions in the D scheme will be
given.
The `generalized' structure functions ${\cal A}$$_{i}(x_{h},Q^{2}_{h})$,
in eq.~(\ref{born}) are:
\begin{eqnarray}
{\cal A}_{1}(x,Q^2)
&=&  2 {\cal F}_{1}^{NC}(x,Q^2)\;,  \nonumber \\
{\cal A}_{2}(x,Q^2)
&=& \frac{1}{yS}
{\cal F}_{2}^{NC}(x,Q^2)\;, \nonumber \\
{\cal A}_{3}(x,Q^2)
&=& { 1 \over 2y S }
{\cal F}_{3}^{NC}(x,Q^2)\;,\nonumber
\end{eqnarray}
with
\begin{eqnarray}
{\cal F}_{1,2}^{NC}(x,Q^2)
&=& F_{1,2}(x,Q^2) + 2 |Q_{e}| \left( v_{e} + \lambda a_e \right)
\chi(Q^2) G_{1,2}(x,Q^2)                    \nonumber          \\
& &+~\left( v_{e}^{2} + a_{e}^{2} + 2 \lambda v_e a_e \right)
\chi^2(Q^2) H_{1,2}(x,Q^2), \nonumber \\
{\cal F}_3^{\rm{NC}}(x,Q^2)
&=& -2sign(Q_e)\left \{|Q_{e}|
\left( a_{e} + \lambda v_e \right) \chi(Q^2) G_{3}(x,Q^2)\right. \nonumber  \\
& & \left.+\left[~2 v_{e} a_e + \lambda \left(v_\e^2 + a_\e^2 \right) \right]
\chi^2(Q^2) H_{3}(x,Q^2)\right \},   \nonumber
\label{f123}
\end{eqnarray}

\noindent
Here structure functions $F_i$, $G_i$ and $H_i$ describe the hadronic tensor
respectively for the $\gamma$, $\gamma-Z$ and $Z$ exchange,
$\lambda=\xi Q_e / |Q_e|$, $\xi$ is the lepton beam polarisation,
$v_{e}$ and $a_{e}$ are the vector and axial-vector
couplings of the lepton to the $Z$~boson:
\begin{eqnarray*}
v_{e}=1-4 |Q_{e}| \sin^{2}\theta_{W}, \hspace{1.cm} a_{e}=1, \nonumber
\end{eqnarray*}
$\theta_{W}$ is the weak mixing angle, $Q_{e}$ is the lepton charge,
$Q_{e}=-1$, and
\begin{eqnarray*}
\chi =  \chi (Q^2) =
{G_\mu \over\sqrt{2}}~{M_{Z}^{2} \over{8\pi\alpha}}~{Q^2 \over
{Q^2+M_{Z}^{2}}},  \nonumber \\
\end{eqnarray*}
with the Fermi constant, $ G_{\mu}= 1.16639\cdot 10^{-5}\,{\rm GeV}^{-2}$.

\medskip
The `radiator' functions  ${\cal S}_i$ and the function ${\cal F}^{IR}$ are:
\begin{eqnarray}
{\cal S}_{1}(y,Q^2,y_h,Q^2_h)
&=& \Biggl\{ \frac{1}{\sqrt{C_{2}}} \left[ \frac{Q^2_h-Q^2}{2}
 +\frac{(Q^2+2m^{2})(Q^2_h-2m^{2})}{Q^2_h-Q^2} \right] \nonumber \\
& &-  m^{2}(Q^2_h-2m^{2})\frac{B_{2}}{C_{2}^{3/2}} \Biggr\}
 - \Biggl\{ S \leftrightarrow - X  \Biggr\} +\frac{1}{\sqrt{A_2}}, \nonumber \\
{\cal S}_{2}(y,Q^2,y_h,Q^2_h)
&=& \Biggl\{\frac{1}{\sqrt{C_{2}}} [ M^{2}(Q^2_h + Q^2)-
XS_h] \nonumber \\
& & +
\frac{1}{(Q^2_h-Q^2)\sqrt{C_{2}}} \Biggl[ Q^2_h[S(S-S_h)+X(X+S_h)
 - 2M^{2}(Q^2_h+2m^2) ]  \nonumber \\
& & + 2m^{2} [(S-S_h)(X+S_h)+SX] \Biggr]
\nonumber \\
& &-2m^{2} \frac{B_{2}}{C_{2}^{3/2}} [S(S-S_h)-M^{2}Q^2_h ] \Biggr\}
- \Biggl\{ S \leftrightarrow -X  \Biggr\}
-\frac{2M^{2}}{\sqrt{A_2}},  \nonumber \\
{\cal S}_{3}(y,Q^2,y_h,Q^2_h) &=&  \Biggl\{\frac{1}{\sqrt{C_{2}}}
\Biggl[\frac{2 Q^2_h ( Q^2_h+2m^{2})(S+X)}{Q^2_h-Q^2}
   - 2XQ^2_h - S_h(Q^2_h+Q^2)\Biggr] \nonumber \\
& & -  2m^{2} Q^2_h\frac{B_{2}}{C_{2}^{3/2}} (2S-S_h)\Biggr\}
+~\Biggl\{ S \leftrightarrow -X   \Biggr\}, \nonumber \\
{\cal F}^{IR}(y,Q^2,y_h,Q^2_h) &=&
         \frac{Q^2+2m^{2}}{Q^2-Q^2_h}
         \Biggl( \frac{1}{\sqrt{C_{1}}}-\frac{1}{\sqrt{C_{2}}} \Biggr)\
         -m^{2} \Biggl(\ \frac{B_{1}}{C_{1}^{3/2}}+
         \frac{B_{2}}{C_{2}^{3/2}} \Biggr)\ ,
\nonumber
\end{eqnarray}
where
\begin{eqnarray}
A_2 &=& \lambda_l \:\equiv \:A_1,  \nonumber       \\
B_2 &=& 2 M^2 Q^2 ( Q^2- Q^2_h ) + X(S_lQ^2_h-S_hQ^2)
    \nonumber \\
    & & +~SQ^2(S_l-S_h)  \:\equiv \:-B_1
    (S \leftrightarrow - X), \nonumber  \\
C_2 &=& [XQ^2_h-Q^2(S-S_h)]^2 + 4m^2
   \left[(S_l-S_h)(S_lQ^2_h-S_hQ^2) \right. \nonumber \\
     & &\left. -~M^2(Q^2_h - Q^2 )^2\right]
    \:\equiv \:C_1 [S \leftrightarrow -X ],
\nonumber
\end{eqnarray}
with
\begin{eqnarray*}
        X=S(1-y) &=& 2ME' \\
             S_h &=& Sy_h \\
\end{eqnarray*}
and the hadron defined variables, $x_h, y_h$ are given by the following
equations
\begin{eqnarray*}
            M_h^2&=&M^2+Sy_h(1-x_h) \\
            Q^2_h&=&Sx_hy_h  \\
\end{eqnarray*}

\medskip\medskip\medskip
\section {Appendix C}
\medskip\medskip
Below we list the relations between the variables used in the MT (Appendix A)
and in the D schemes (Appendix B):

\begin{eqnarray*}
 \mbox{\rm Dubna scheme} &    & \mbox{Mo and Tsai scheme}\\
        E &    & E_s \\
       E' &    & E_p \\
   S_l=Sy &    & 2M\nu \\
    Q^2_h &    & -q^2 \\
\lambda_l=S_l^2 + 4M^2Q^2  &    &  \left(2M\mid\overline u\mid\right)^2 \\
B_2\over{2\lambda_l}&     & a \\
C_2\over{4\lambda_l}&     &- x^2 \\
M_h &    & M_j \\
{2M\over{\left(\lambda_l\right)^{1/2}}}\left(E - E' +
M\left({Q^2_h-Q^2}\over{Q^2_h-S_l}\right)\right)&& \cos\theta_k \\
{dQ^2_h}\over{Q^4_h\lambda_l^{1/2}}&&{\D\left(\cos\theta_k\right)\omega}\over
{q^4\left(u_0 - \mid\overline u\mid \cos\theta_k\right)}  \\
\end{eqnarray*}

\medskip \medskip \medskip
\noindent
{\Large {\bf {Figure Captions}}}
\medskip \medskip
\begin{enumerate}
\item Feynman diagrams for the deep inelastic scattering in the one photon
exchange approximation (a) and the lowest order radiative processes: real
photon bremsstrahlung from the charged lepton (b), vertex correction (c),
vacuum polarisation (d), double photon exchange (e), hadron current
corrections (f,g). In the MT scheme evaluated were diagrams b--d.

\item Range of kinematical variables from which the radiative tails contribute
to the cross section measured at the point A($Q^2,\nu$).

\item Double photon exchange (a) and hadron current corrections (b,c) in the
Dubna scheme.

\item Infrared cut--off parameter $\Delta$ dependence of the FERRAD35 results
obtained for the 280 GeV muon -- proton scattering.
Radiative correction factor $\eta$ is defined in eq.(\ref{eta}).

\item Radiative correction factor $\eta$ calculated in FERRAD35 (open symbols)
 and TERAD86 (closed symbols) for the muon -- proton scattering at 280 GeV.

\item Ratio of the radiative correction factors $\eta$ calculated in FERRAD35
($\eta_F$) and TERAD86 ($\eta_T$) for the muon -- proton scattering at 280~GeV
(closed symbols). The open symbols give the $\eta_F/\eta_T$ ratio when only
$e^+e^-$ and $\mu^+\mu^-$ contribute to the vacuum polarisation (fig.1d)
in FERRAD35.

\item Definition of kinematic variables describing the hard photon emission
in MT, cf. Appendix A (from \cite{Drees}).

\item The coordinate system used in the integration over the solid
angle of the photon in the formula A.24 in \cite{MTb}, quoted in Appendix A
(from \cite{MTb}).

\end{enumerate}

\end{document}